\documentclass{article}

\def\xxinput#1{\input#1}

\xxinput{vsolj02.sty}

\usepackage[dvipdfmx]{graphicx}
\usepackage[comma,colon]{natbib}

\usepackage[OT2,T1]{fontenc}

\def\cite{\citealt}
\setcitestyle{aysep={}}

\newcounter{author}
\def\altaffilmark#1{$^{#1}$}
\def\altaffiltext#1{$^{#1}$\,}

\def\authorcount#1#2{{\refstepcounter{author}\label{#1}
                     \altaffiltext{\ref{#1}}{#2}}}

\begin{document}

\begin{center}

\title{ASASSN-15cm: an SU UMa star with an orbital period of 5.0 hours}

\author{
        Taichi~Kato\altaffilmark{\ref{affil:Kyoto}}}
\email{tkato@kusastro.kyoto-u.ac.jp}

\authorcount{affil:Kyoto}{
     Department of Astronomy, Kyoto University, Sakyo-ku,
     Kyoto 606-8502, Japan}
\end{center}

\begin{abstract}
\xxinput{abst.inc}
\end{abstract}

\section{Introduction}

   ASASSN-15cm was discovered as a dwarf nova by
the All-Sky Automated Survey for
Supernovae (ASAS-SN: \cite{ASASSN}) at $V$=15.8 on
2015 January 31.  The ASAS-SN team noted that
the Catalina Real-time Transient Survey
(CRTS, \cite{dra08atel1734})\footnote{
   $<$http://nesssi.cacr.caltech.edu/catalina/$>$.
} had detected previous outbursts.
\citet{tho16CVs} performed spectroscopic and photometric
studies of this object and obtained an orbital period
($P_{\rm orb}$) of 0.208466(2)~d using the CRTS data.
Spectroscopy by \citet{tho16CVs} indicated a secondary of
a spectral type of K2.5$\pm$2.5, hot and luminous for
a cataclysmic variable (CV) with this $P_{\rm orb}$
Based on the long $P_{\rm orb}$, the object was implicitly
assumed to be an SS Cyg star and no special attention
has been paid to search for an SU UMa-type signature.
[For general information of CVs and subclasses,
see e.g., \citet{war95book}.  For a review of dwarf novae,
including SU UMa stars, see e.g., \citet{osa96review}].

\section{Data}

   I used the Asteroid Terrestrial-impact Last Alert System
(ATLAS: \cite{ATLAS}) forced photometry \citep{shi21ALTASforced}
and the Zwicky Transient Facility (ZTF: \cite{ZTF})\footnote{
   The ZTF data can be obtained from IRSA
$<$https://irsa.ipac.caltech.edu/Missions/ztf.html$>$
using the interface
$<$https://irsa.ipac.caltech.edu/docs/program\_interface/ztf\_api.html$>$
or using a wrapper of the above IRSA API
$<$https://github.com/MickaelRigault/ztfquery$>$.
} data together with ASAS-SN Sky Patrol data \citep{koc17ASASSNLC}.

\section{Results and Discussions}

\begin{figure*}
\begin{center}
\includegraphics[width=16cm]{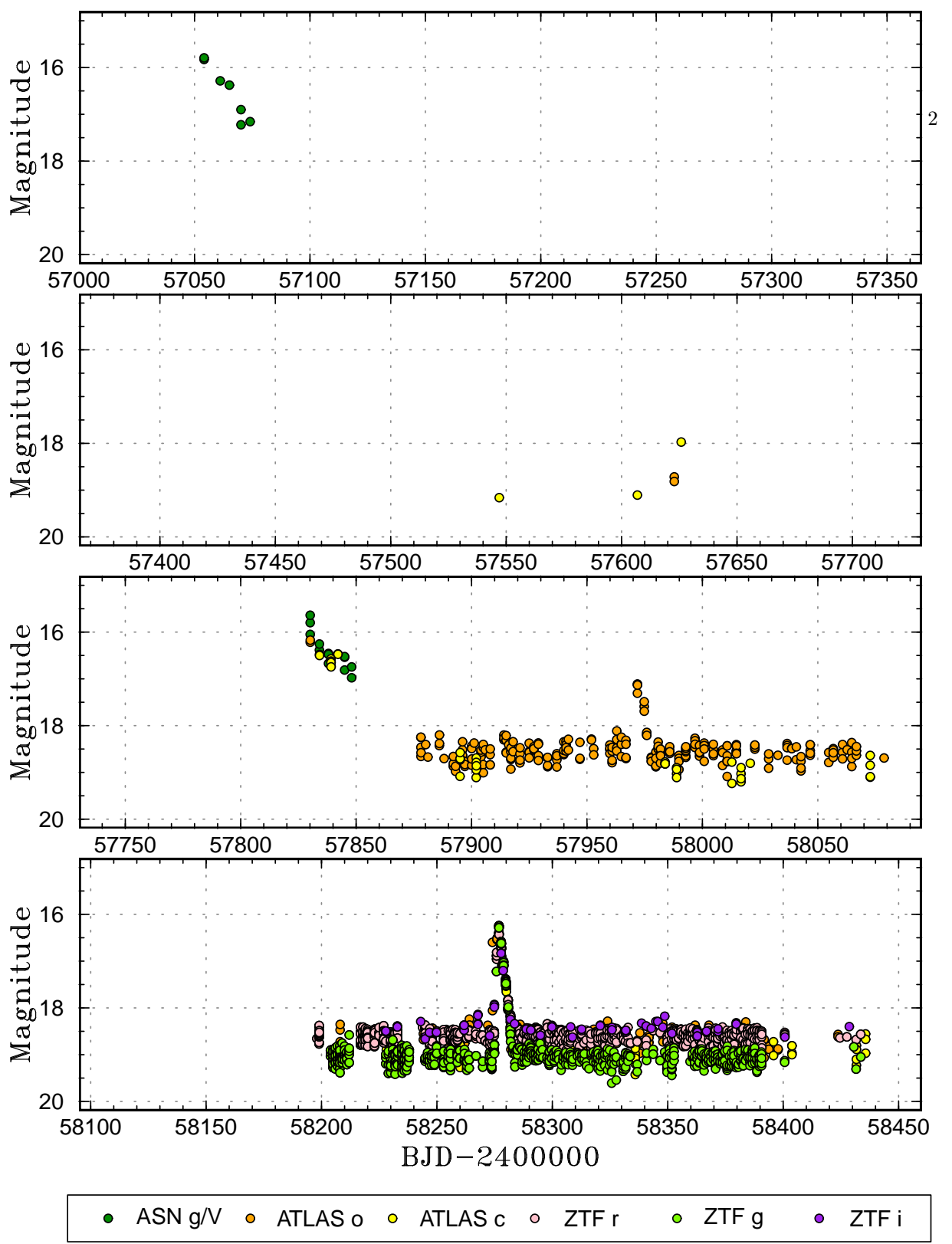}
\caption{
   Light curve of ASASSN-15cm in 2015--2018.
}
\label{fig:lc1}
\end{center}
\end{figure*}

\begin{figure*}
\begin{center}
\includegraphics[width=16cm]{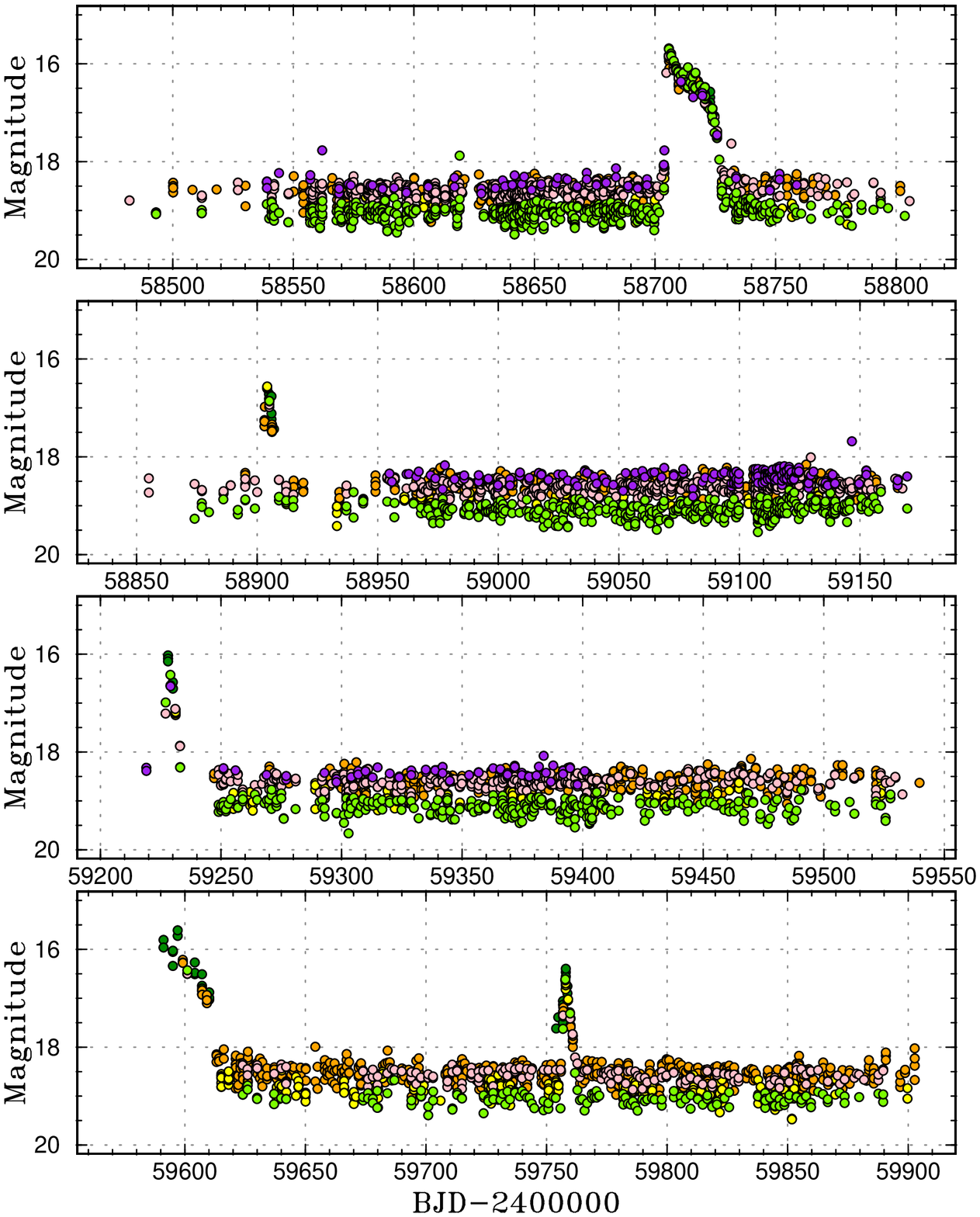}
\caption{
   Light curve of ASASSN-15cm in 2019--2022.
The symbols are the same as in figure \ref{fig:lc1}.
}
\label{fig:lc2}
\end{center}
\end{figure*}

\subsection{Superoutbursts and superhumps}\label{sec:SH}

   In figures \ref{fig:lc1} and \ref{fig:lc2}, I show
the light curve of ASASSN-15cm in the 2015--2022 seasons.
ASAS-SN data did not reach 18.0 mag and quiescent
parts was not recorded.  In all these surveys, there were false
bright detections and they were removed by comparing
with the data obtained on the same nights by the same or
other surveys.  Upper limit observations were
not plotted to avoid the figures to be too crowded.

\begin{figure*}
\begin{center}
\includegraphics[width=16cm]{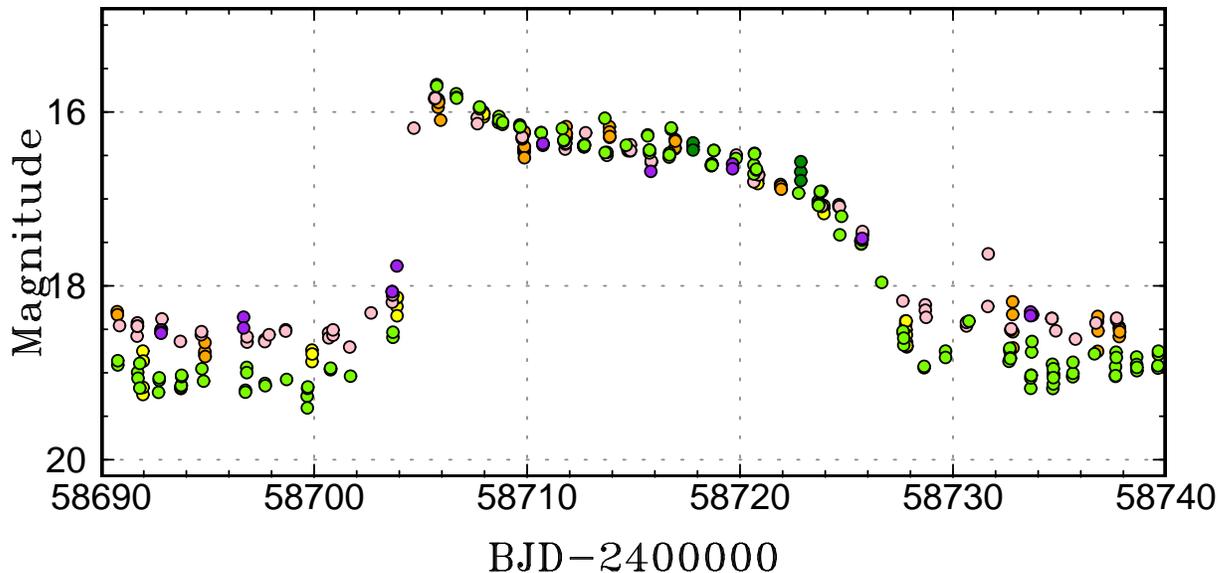}
\caption{
   Enlargement of the light curve of the 2019 superoutburst.
   The symbols are the same as in figure \ref{fig:lc1}.
}
\label{fig:lcso}
\end{center}
\end{figure*}

\begin{figure*}
\begin{center}
\includegraphics[width=14cm]{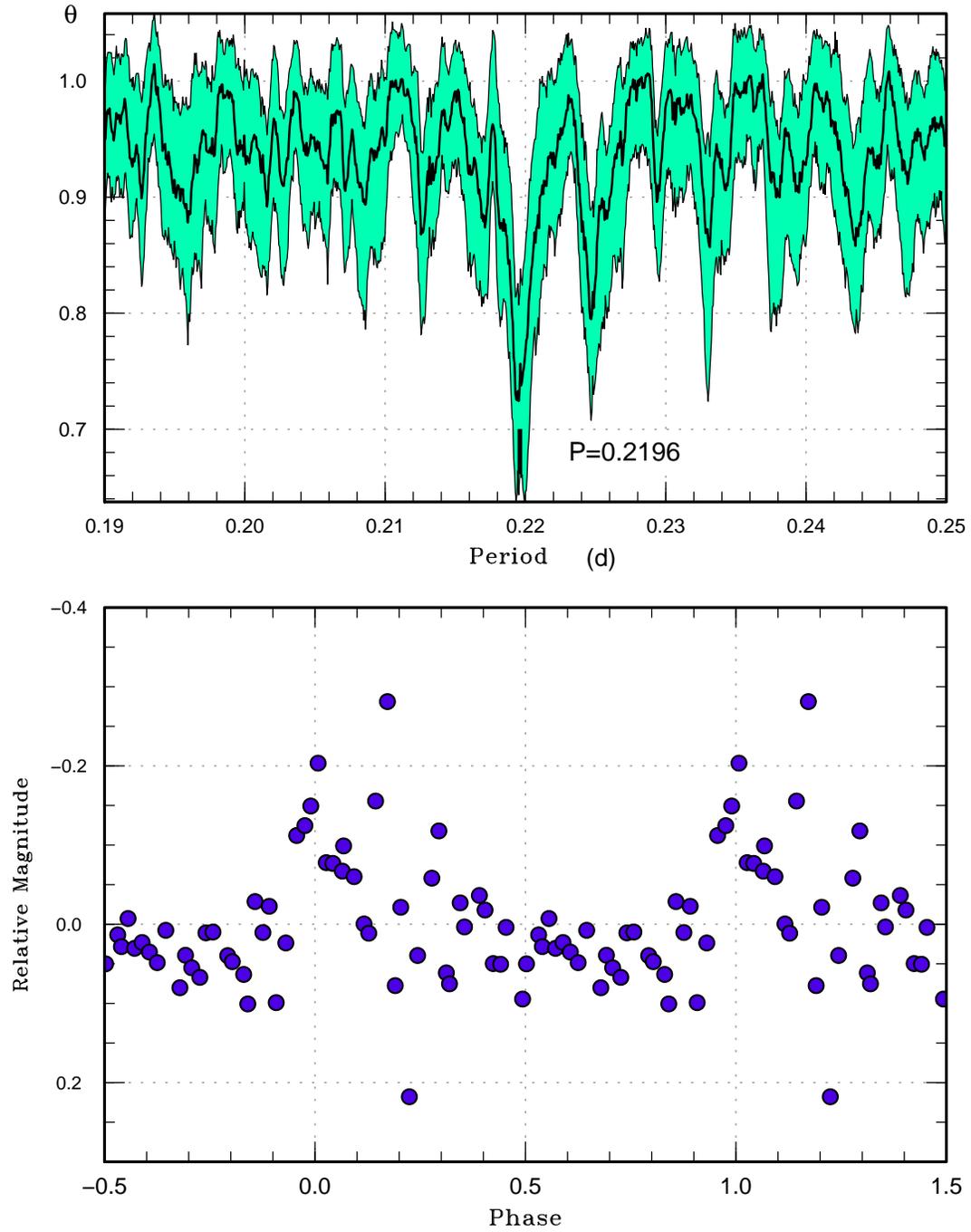}
\caption{
   Superhumps of ASASSN-15cm in 2019 (interval BJD 2458705.6--2458726.7).
   (Upper): PDM analysis.  The bootstrap result using
   randomly contain 50\% of observations is shown as
   a form of 90\% confidence intervals in the resultant 
   $\theta$ statistics.
   (Lower): Phase plot.
}
\label{fig:shpdm}
\end{center}
\end{figure*}

   The initial outburst in 2015 (first panel of figure
\ref{fig:lc1}) corresponds to the initial announcement of
the outburst detection by the ASAS-SN team.  I noticed
that a long outburst in 2019 (first panel of figure
\ref{fig:lc2}) had a shape similar to an SU UMa-type
superoutburst.  This outburst was composed of a linearly
fading initial part and a subsequent part with a scatter
in the light curve (figure \ref{fig:lcso}).
Using these data after removing the trend by
locally-weighted polynomial regression (LOWESS: \cite{LOWESS}),
a phase dispersion minimization (PDM, \cite{PDM}) analysis
yielded a strong superhump signal (figure \ref{fig:shpdm})
having a period of 0.2196(1)~d with its error determined by
the methods of \citet{fer89error} and \citet{Pdot2}.
This period analysis had been proven to be useful
for sparse, but with multiple samples in the same night, data
such as ATLAS observations \citep{kat23bocet}.
Note that the entire part was analyzed and possible
variation in the superhump period (e.g. \cite{Pdot}) may have
affected the result.  There was no hint of the orbital
modulation during the superoutburst.

   ASASSN-15cm showed long outbursts (likely superoutbursts)
relatively regularly: in 2015 (BJD 2457054), in 2017 (BJD 2457830),
in 2019 (BJD 2458706) and in 2022 (BJD 2459591), where BJD values
refer to the peaks of the outbursts
(figures \ref{fig:lc1}, \ref{fig:lc2}).
These long outbursts could be expressed by a period (supercycle)
of 849(18)~d within errors of 41~d.  This result suggests
that superbursts in ASASSN-15cm occur very regularly, and
the 2019 superoutburst was unlikely an exceptional phenomenon.
The next superoutburst is expected to occur in the early
half of 2024 (around BJD 2460417) and coordinated detailed
observations are desirable.
The supercycle of 849~d is fairly long and this implies
a low mass-transfer rate.

   The object also showed normal outbursts.  These outbursts
were relatively slowly rising and could be inside-out
outbursts.  The small number of normal outbursts also
supports the low mass-transfer rate \citep{ich94cycle,osa96review}.

\subsection{Orbital period and variation in quiescence}

\begin{figure*}
\begin{center}
\includegraphics[width=14cm]{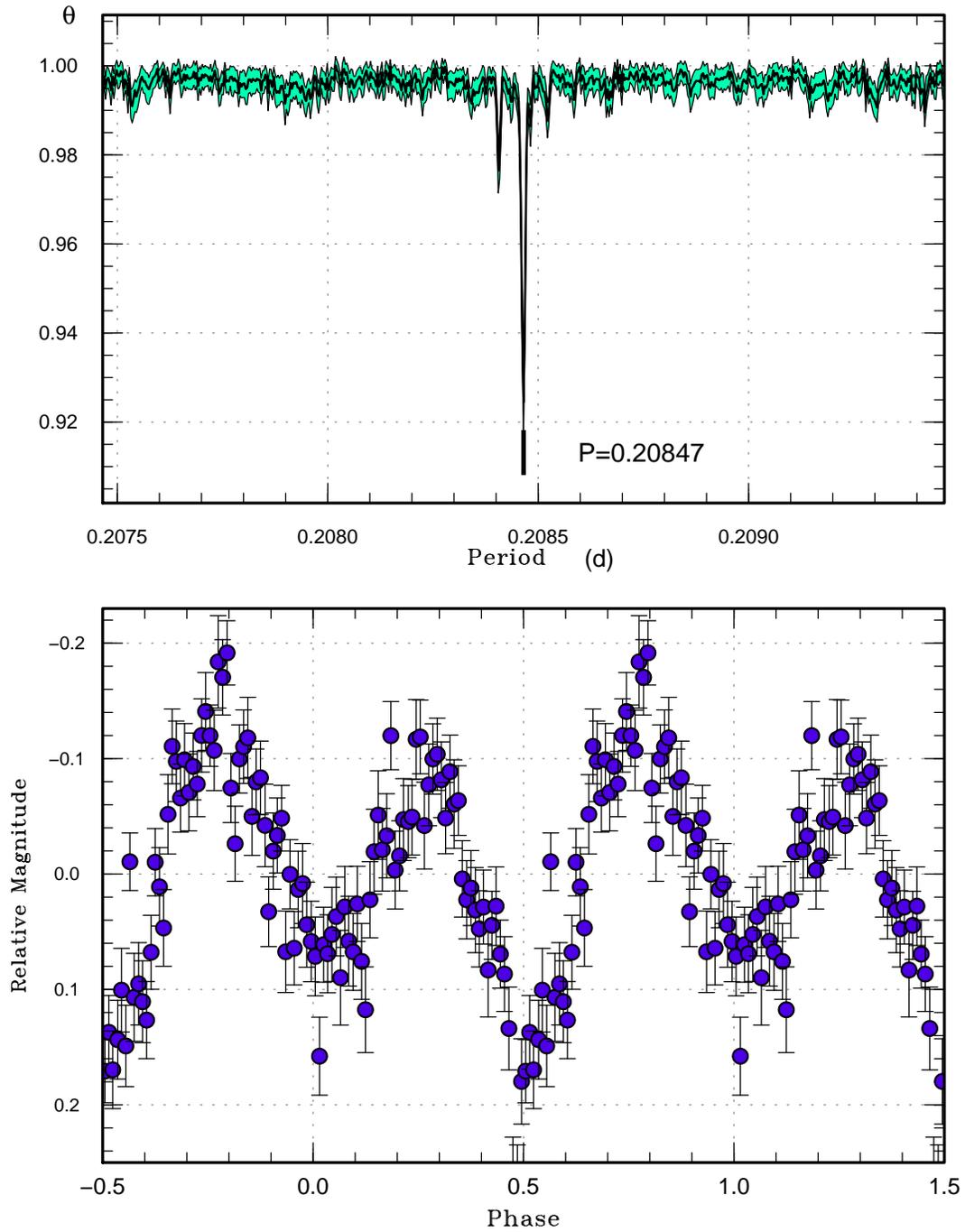}
\caption{
   Orbital profile of ASASSN-15cm in quiescence using
   ATLAS and ZTF data.
   (Upper): PDM analysis.
   (Lower): Phase plot.  Prominent ellipsoidal variations are
   present.  The zero phase was chosen to be
   BJD 2457553.8461 to match figure 15 in \citet{tho16CVs}.
   This zero phase corresponds to the inferior conjunction
   of the secondary.
}
\label{fig:porbpdm}
\end{center}
\end{figure*}

   Using ATLAS and ZTF data (all bands combined), a PDM
analysis of the quiescent part yielded an orbital period
of 0.2084652(3)~d (figure \ref{fig:porbpdm}),
in very good agreement with the value by \citet{tho16CVs}.
The quiescent light curve consists of prominent ellipsoidal
variations, which are expected from the hot and luminous
secondary \citep{tho16CVs}.  The zero phase was chosen to be
BJD 2457553.8461 to match figure 15 in \citep{tho16CVs}.
This zero phase corresponds to the inferior conjunction
of the secondary.  Multicolor phase-averaged light curves
are shown in figure \ref{fig:quilc}.  The maximum at
phase 0.7--0.8 (possibly corresponding to the hot spot, but
the O'Connell effect cannot be excluded)
becomes brighter in the $r$ and $g$ bands than in the $i$ band.

\subsection{Mass ratio and orbital parameters}

   The observed fractional superhump excess
$\epsilon \equiv P_{\rm SH}/P_{\rm orb}-1$, where
$P_{\rm SH}$ represents the superhump period, is 0.053.
Assuming that $P_{\rm SH}$ represents the period of
stage B superhumps
[for superhumps and stages, see \citet{Pdot}],
the semi-empirically derived $q=M_2/M_1$ is 0.22
(table 4 in \cite{kat22stageA}).
Although I do not explicitly give an error estimate,
the error is expected to be an order of 0.01
[see \citet{kat22stageA} for the derivation process].
This appears to fit the observations fairly
well since $q<$0.25--0.33 is required to show
superhumps during outbursts
\citep{whi88tidal,mur00SHintermediateq}.
The regularity of superoutbursts (subsection \ref{sec:SH})
provides a support to the $q$ value not too close to
the borderline, i.e. not close to 0.3 or larger.

   This $q$ is very different from $q \sim$0.6 suggested in
\citet{tho16CVs}.\footnote{
   The modest lower limit of $q$=0.1 by \citet{tho16CVs},
   however, turned out to be an excellent choice after
   the identification of this system as an SU UMa star.
}  This constraint by \citet{tho16CVs} on $q$ was based on
a combination of the observed $K_2$ velocity and the constraint
on the inclination allowing a large amplitude of
ellipsoidal variations, which could not be comfortably
modeled without introducing artificially increasing
gravity darkening and decreasing
the disk contribution excessively \citep{tho16CVs}.
The second assumption (negligible contribution from the disk)
required for modeling at least appears to be satisfied
considering the low mass-transfer rate inferred from
the long supercycle (subsection \ref{sec:SH}).

\begin{figure*}
\begin{center}
\includegraphics[width=14cm]{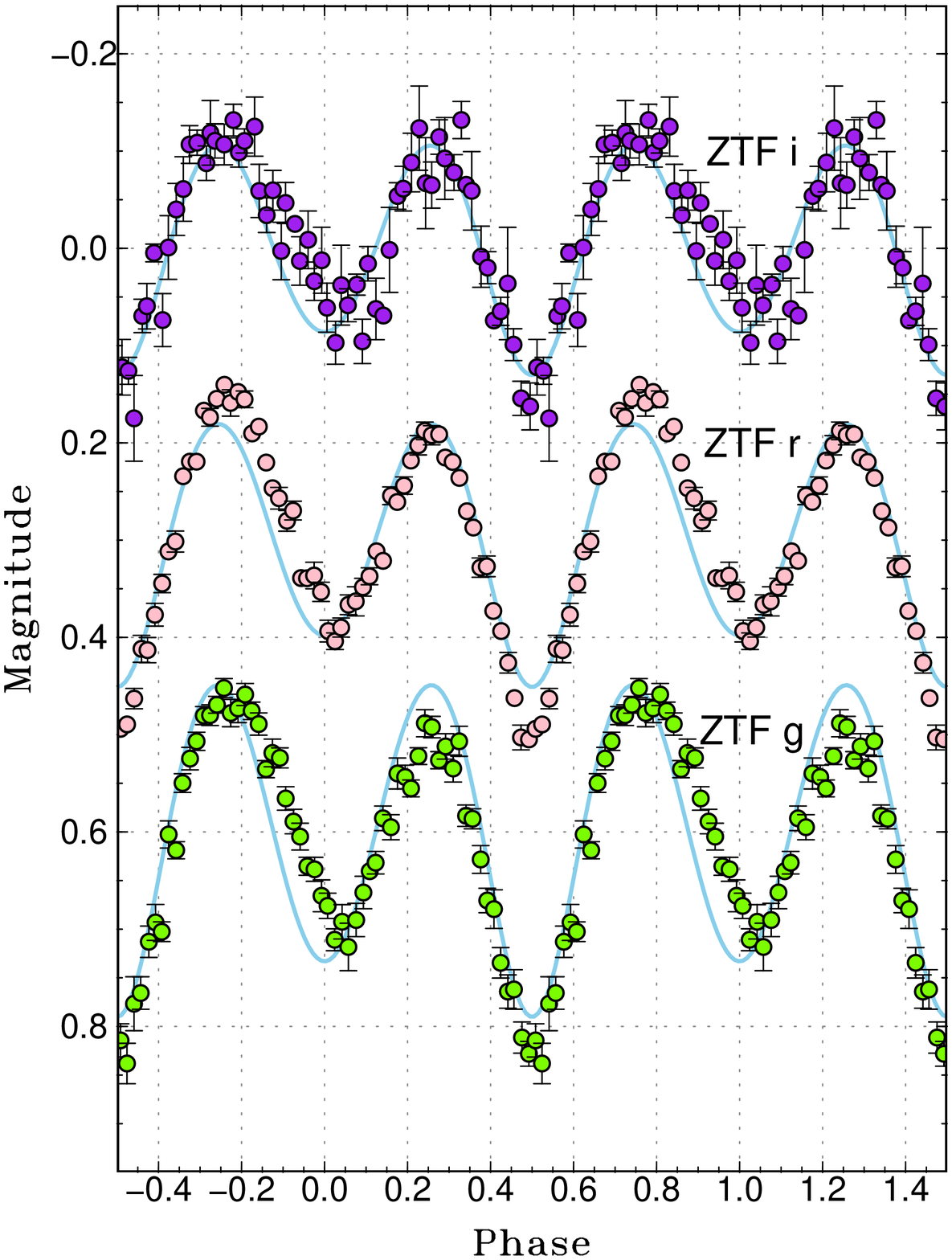}
\caption{
   Orbital profile of ASASSN-15cm in quiescence using
   multicolor ZTF data.  The zero phase is defined as the same
   as in figure \ref{fig:porbpdm}.  The plots were shifted by 0.3~mag
   between different bands.
   The solid curves represent ellipsoidal variations
   expected for $q$=0.22 and $i$=55$^\circ$ (see text).
}
\label{fig:quilc}
\end{center}
\end{figure*}

   Using Ellipsoidal Modulation Light Curve Generator by
M.~Uemura (2006), which was based on
\citet{oro97j1655,oro97j1655erratum}, I obtained ellipsoidal
variations similar to the ZTF light curves
in figure \ref{fig:quilc}.  I used $T_{\rm eff}$=5000~K,
$\log g$=4.5 and gravitational darkening
and limb-darkening coefficients given in \citet{cla11limbdark}
(solar metallicity and a microturbulent velocity of 2~km s$^{-1}$
as typical values were used, since I do not have information
for them; however, they have little effect on the result).
With these parameters, the primary with a mass of 0.7~$M_\odot$
can reproduce the observed $K_2$=229 km s$^{-1}$.

\subsection{SU UMa stars above or in the period gap}

   SU UMa stars above the CV period gap are relatively
rare, since CVs with long periods have larger $q$ values,
which make the 3:1 resonance in the disk difficult to
achieve.  The present finding in ASASSN-15cm adds another
example of SU UMa stars above or in the period gap containing
a secondary with an evolved core.  The known such systems
include OT J002656.6$+$284933 (=CSS101212:002657$+$284933,
$P_{\rm SH}$=0.13225~d, \cite{kat17j0026}),
ASASSN-18aan ($P_{\rm orb}$=0.149454~d, \cite{wak21asassn18aan})
and likely ASASSN-19ax ($P_{\rm SH}$=0.1000~d, \cite{kat21asassn19ax})
following the identification of QZ Ser (but near the lower
border of the period gap with $P_{\rm orb}$=0.08316~d)
as an SU UMa star with an anomalously
hot and luminous secondary \citep{tho02qzser}.
V363 Lyr ($P_{\rm orb}$=0.185723~d, \cite{kat21v363lyr})
might be another example, although the superhump-like signal
recorded in this system may have not been the same as
in other SU UMa stars.
Another group of SU UMa stars above the period gap,
but apparently lacking evidence of a secondary with
an evolved core, contains BO Cet
($P_{\rm orb}$=0.139835~d, \cite{kat21bocet,kat23bocet}) and
possibly MisV1448 ($P_{\rm SH}$=0.2275~d, vsnet-alert 24912,\footnote{
  $<$http://ooruri.kusastro.kyoto-u.ac.jp/mailarchive/vsnet-alert/24912$>$.
} Kojiguchi et al. in preparation).
The status of ASASSN-14ho
($P_{\rm orb}$=0.24315~d, \cite{kat20asassn14ho}) still requires
observational confirmation.  Lists of long-$P_{\rm orb}$
SU UMa stars or CVs with an anomalously warm secondary can be
found in \citet{wak21asassn18aan} and
\citet{kat21asassn19ax}.  A search for superhumps in
long-$P_{\rm orb}$ systems with a secondary with an evolved core
would be an interesting challenge.

\section*{Acknowledgements}

This work was supported by JSPS KAKENHI Grant Number 21K03616.

I am grateful to Makoto Uemura for sharing the code of
Ellipsoidal Modulation Light Curve Generator, Naoto Kojiguchi
for helping downloading the ZTF data and
the ZTF, ATLAS and ASAS-SN teams for making their data
available to the public.

Based on observations obtained with the Samuel Oschin 48-inch
Telescope at the Palomar Observatory as part of
the Zwicky Transient Facility project. ZTF is supported by
the National Science Foundation under Grant No. AST-1440341
and a collaboration including Caltech, IPAC, 
the Weizmann Institute for Science, the Oskar Klein Center
at Stockholm University, the University of Maryland,
the University of Washington, Deutsches Elektronen-Synchrotron
and Humboldt University, Los Alamos National Laboratories, 
the TANGO Consortium of Taiwan, the University of 
Wisconsin at Milwaukee, and Lawrence Berkeley National Laboratories.
Operations are conducted by COO, IPAC, and UW.

The ztfquery code was funded by the European Research Council
(ERC) under the European Union's Horizon 2020 research and 
innovation programme (grant agreement n$^{\circ}$759194
-- USNAC, PI: Rigault).

This work has made use of data from the Asteroid Terrestrial-impact
Last Alert System (ATLAS) project. The Asteroid Terrestrial-impact
Last Alert System (ATLAS) project is primarily funded to search for
near earth asteroids through NASA grants NN12AR55G, 80NSSC18K0284,
and 80NSSC18K1575; byproducts of the NEO search include images and
catalogs from the survey area. This work was partially funded by
Kepler/K2 grant J1944/80NSSC19K0112 and HST GO-15889, and STFC
grants ST/T000198/1 and ST/S006109/1. The ATLAS science products
have been made possible through the contributions of the University
of Hawaii Institute for Astronomy, the Queen's University Belfast, 
the Space Telescope Science Institute, the South African Astronomical
Observatory, and The Millennium Institute of Astrophysics (MAS), Chile.

\section*{List of objects in this paper}
\xxinput{objlist.inc}

\section*{References}

We provide two forms of the references section (for ADS
and as published) so that the references can be easily
incorporated into ADS.

\newcommand{\noop}[1]{}\newcommand{\hyphalt}{-}

\renewcommand\refname{\textbf{References (for ADS)}}

\xxinput{asn15cmaph.bbl}

\renewcommand\refname{\textbf{References (as published)}}

\xxinput{asn15cm.bbl.vsolj}


\begin{thebibliography}{}

\bibitem[{Claret} and {Bloemen}(2011)]{cla11limbdark}
  {Claret}, A., \& {Bloemen}, S.\ 2011, A\&A, 529, A75
  (https://doi.org/10.1051/0004-6361/201116451)

\bibitem[{Cleveland}(1979)]{LOWESS}
  {Cleveland}, W.~S.\ 1979, J. Amer. Statist. Assoc., 74, 829
  (https://doi.org/10.2307/2286407)

\bibitem[{Drake} et~al.(2008)]{dra08atel1734}
  {Drake}, A.~J., {et~al.}\ 2008, Astron.\ Telegram, 1734

\bibitem[{Fernie}(1989)]{fer89error}
  {Fernie}, J.~D.\ 1989, PASP, 101, 225 (https://doi.org/10.1086/132426)

\bibitem[Ichikawa and Osaki(1994)]{ich94cycle}
  Ichikawa, S., \& Osaki, Y.\ 1994, in Theory of Accretion Disks-2, ed. W.~J.
  Duschl, J. Frank, F. Meyer, E. Meyer-Hofmeister, \& W.~M. Tscharnuter
  (Dordrecht: Kluwer Academic Publishers),  p.~169

\bibitem[{Kato}(2020)]{kat20asassn14ho}
  {Kato}, T.\ 2020, PASJ, 72, L2 (arXiv:1911.08093)

\bibitem[{Kato}(2021)]{kat21v363lyr}
  {Kato}, T.\ 2021, VSOLJ\ Variable\ Star\ Bull., 85, (arXiv:2111.07237)

\bibitem[{Kato}(2022)]{kat22stageA}
  {Kato}, T.\ 2022, VSOLJ\ Variable\ Star\ Bull., 89, (arXiv:2201.02945)

\bibitem[{Kato}(2023)]{kat23bocet}
  {Kato}, T.\ 2023, VSOLJ\ Variable\ Star\ Bull., 106, (arXiv:2302.02593)

\bibitem[{Kato} et~al.(2009)]{Pdot}
  {Kato}, T., {et~al.}\ 2009, PASJ, 61, S395 (arXiv:0905.1757)

\bibitem[{Kato} et~al.(2021a)]{kat21asassn19ax}
  {Kato}, T., {et~al.}\ 2021a, VSOLJ\ Variable\ Star\ Bull., 84,
  (arXiv:2111.01304)

\bibitem[{Kato} et~al.(2010)]{Pdot2}
  {Kato}, T., {et~al.}\ 2010, PASJ, 62, 1525 (arXiv:1009.5444)

\bibitem[{Kato} et~al.(2021b)]{kat21bocet}
  {Kato}, T., {et~al.}\ 2021b, PASJ, 73, 1280 (arXiv:2106.15028)

\bibitem[{Kato} et~al.(2017)]{kat17j0026}
  {Kato}, T., {et~al.}\ 2017, PASJ, 69, L4 (arXiv:1703.00650)

\bibitem[{Kochanek} et~al.(2017)]{koc17ASASSNLC}
  {Kochanek}, C.~S., {et~al.}\ 2017, PASP, 129, 104502 (arXiv:1706.07060)

\bibitem[{Masci} et~al.(2019)]{ZTF}
  {Masci}, F.-J., {et~al.}\ 2019, PASP, 131, 018003 (arXiv:1902.01872)

\bibitem[{Murray} et~al.(2000)]{mur00SHintermediateq}
  {Murray}, J., {Warner}, B., \& {Wickramasinghe}, D.\ 2000, New\ Astron.\
  Rev., 44, 51

\bibitem[{Orosz} and {Bailyn}(1997a)]{oro97j1655}
  {Orosz}, J.~A., \& {Bailyn}, C.~D.\ 1997a, ApJ, 477, 876
  (arXiv:astro-ph/9610211)

\bibitem[{Orosz} and {Bailyn}(1997b)]{oro97j1655erratum}
  {Orosz}, J.~A., \& {Bailyn}, C.~D.\ 1997b, ApJ, 482, 1086
  (https://doi.org/10.1086/318114)

\bibitem[{Osaki}(1996)]{osa96review}
  {Osaki}, Y.\ 1996, PASP, 108, 39 (https://doi.org/10.1086/133689)

\bibitem[{Shappee} et~al.(2014)]{ASASSN}
  {Shappee}, B.~J., {et~al.}\ 2014, ApJ, 788, 48 (arXiv:1310.2241)

\bibitem[{Shingles} et~al.(2021)]{shi21ALTASforced}
  {Shingles}, L., {et~al.}\ 2021, Transient Name Server AstroNote, 7, 1

\bibitem[{Stellingwerf}(1978)]{PDM}
  {Stellingwerf}, R.~F.\ 1978, ApJ, 224, 953 (https://doi.org/10.1086/156444)

\bibitem[{Thorstensen} et~al.(2002)]{tho02qzser}
  {Thorstensen}, J.~R., {Fenton}, W.~H., {Patterson}, J.~O., {Kemp}, J.,
  {Halpern}, J., \& {Baraffe}, I.\ 2002, PASP, 114, 1117
  (arXiv:astro-ph/0206435)

\bibitem[{Thorstensen} et~al.(2016)]{tho16CVs}
  {Thorstensen}, J.~R., {Alper}, E.~H., \& {Weil}, K.~E.\ 2016, AJ, 152, 226
  (arXiv:1609.02215)

\bibitem[{Tonry} et~al.(2018)]{ATLAS}
  {Tonry}, J.~L., {et~al.}\ 2018, PASP, 130, 064505 (arXiv:1802.00879)

\bibitem[{Wakamatsu} et~al.(2021)]{wak21asassn18aan}
  {Wakamatsu}, Y., {et~al.}\ 2021, PASJ, 73, 1209 (arXiv:2102.04104)

\bibitem[{Warner}(1995)]{war95book}
  {Warner}, B.\ 1995, Cataclysmic Variable Stars
 (Cambridge: Cambridge University Press)

\bibitem[{Whitehurst}(1988)]{whi88tidal}
  {Whitehurst}, R.\ 1988, MNRAS, 232, 35
  (https://doi.org/10.1093/mnras/232.1.35)

\end{thebibliography}
\end{document}